\documentstyle[aps,psfig,floats,epsf,twocolumn,prb]{revtex}

\begin{document}
\tighten
\draft
\twocolumn[
\hsize\textwidth\columnwidth\hsize\csname @twocolumnfalse\endcsname  

\title{Raman scattering through a metal-insulator transition}
\author{J.K. Freericks$^*$ and T.P. Devereaux$^\dagger$}
\address{$^*$Department of Physics, Georgetown University, Washington, DC
20057, U.S.A.}
\address{$^\dagger$Department of Physics, University of Waterloo, Waterloo, ON
N2G 1Y2, Canada}
\date{\today}
\maketitle

\widetext
\begin{abstract}
The exact solution for nonresonant $A_{\rm 1g}$ and $B_{\rm 1g}$
Raman scattering is presented for the simplest model that
has a correlated metal-insulator transition---the Falicov-Kimball model,
by employing dynamical mean field theory.
In the general case,
the $A_{\rm 1g}$ response includes nonresonant, 
resonant, and mixed contributions, the $B_{\rm 1g}$ response includes
nonresonant and resonant contributions (we prove the Shastry-Shraiman 
relation for the nonresonant $B_{\rm 1g}$ response) while the $B_{\rm 2g}$
response is purely resonant.  Three main features are seen in 
the nonresonant $B_{\rm 1g}$ channel: (i) the rapid appearance
of low-energy spectral weight at the expense of higher-energy weight; (b)
the frequency range for this low-energy spectral weight is much larger than
the onset temperature, where the response first appears; and (iii) the occurrence
of an isosbestic point, which is a characteristic frequency where the
Raman response is independent of temperature for low temperatures. Vertex
corrections renormalize away all of these anomalous features in the nonresonant
$A_{\rm 1g}$ channel. The calculated results compare favorably to the Raman
response of a number of correlated systems on the insulating side of the
quantum-critical point (ranging from Kondo insulators, to mixed-valence
materials,
to underdoped
high-temperature superconductors).  We also show why the nonresonant
$B_{\rm 1g}$ Raman
response is ``universal'' on the insulating side of the metal-insulator 
transition.
\end{abstract}
]
\narrowtext
\section{Introduction}

Raman scattering has been an important
experimental tool for studying lattice dynamics
for over four decades (since the advent of the laser).  More recently, it
has been applied to study the scattering of electrons in metals, insulators,
semiconductors, and superconductors. Via light's coupling to the electron's 
charge, inelastic light scattering reveals symmetry selective properties of the
electron dynamics over a wide range of energy scales and temperatures.  It
is similar to the optical conductivity, which involves
elastic scattering of light by electron-hole pairs; but Raman scattering 
provides additional information,
since it is able to isolate different symmetry channels
by selectively polarizing the incident light and 
measuring the reflected light with a polarized detector.  Three principle
symmetries are usually examined: (i) $A_{\rm 1g}$ which has the full symmetry of
the lattice (i.e. is $s$-like); (ii) $B_{\rm 1g}$ which is a $d$-like symmetry
(that probes the Brillouin-zone axes); and (iii) $B_{\rm 2g}$ which is another
$d$-like symmetry (that probes the Brillouin-zone diagonals).
In 1990-91, Shastry and Shraiman proposed a simple relation\cite{ss} that 
connects the nonresonant
Raman response to the optical conductivity.  We prove 
the Shastry-Shraiman relation here for the $B_{1g}$ channel
in the large-dimensional limit.

Strongly correlated systems as disparate 
as mixed-valence compounds\cite{SLC1} (such as SmB$_{6}$), 
Kondo-insulators\cite{SLC2}
(such as FeSi), and the underdoped cuprate high temperature
superconductors\cite{irwin,hackl,uiuc}, 
show temperature-dependent $B_{\rm 1g}$
Raman spectra that are both remarkably similar and quite anomalous.
This ``universality''
suggests that there is a common mechanism governing the electronic
transport in correlated insulators. As these materials are cooled, they all 
show a pile up of spectral weight for moderate photon energy losses
with a simultaneous reduction of the low-frequency spectral weight.
This spectral weight transfer is slow at high temperatures and then
rapidly increases as the temperature is lowered towards a putative
quantum critical point (corresponding to a metal-insulator transition). 
In addition, the Raman spectral range
is divided into two regions: one where the response decreases as $T$
is lowered and one where the response increases.  These regions are
separated by a so-called isosbestic point, which is defined to be the
characteristic frequency where the Raman response is independent of
temperature. Finally, it is often observed that the range of frequency
where the Raman response is reduced as $T$ is lowered is an order
of magnitude (or more) larger than the temperature at which the
low-frequency spectral weight disappears. These anomalous features are
not typically seen in either the $A_{\rm 1g}$ or the $B_{\rm 2g}$ channels.

Theory has lagged behind experiment for electronic Raman scattering in
strongly correlated materials.
While theories that describe Raman scattering in
weakly correlated (Fermi-liquid) metals\cite{metals} or in band
insulators\cite{ins} have been known for some time, it is only recently
that a theory that describes materials near the metal-insulator transition
has been developed\cite{raman_us}.  This theoretical treatment involves
applying the dynamical mean-field theory to the simplest many-body system
that has a quantum-critical point---the spinless Falicov-Kimball 
model\cite{falicov_kimball}.  We choose this model
as our canonical model for Raman scattering because
it can be solved exactly and the results are universal for the nonresonant
$B_{\rm 1g}$ channel on the insulating
side of the transition (we do not choose it because we believe the 
Falicov-Kimball model is the physically relevant model for
all correlated experimental systems---it isn't).  Other 
work has been performed in  the so-called iterated-perturbation-theory
approximation to the Hubbard model, where the nonresonant $B_{\rm 1g}$
Raman response is determined\cite{craco}.

The Hamiltonian
contains two types of electrons: itinerant band electrons and localized
(d or f) electrons.  The band electrons can hop between nearest neighbors
[with hopping integral $t^*/(2\sqrt{d})$ on a $d$-dimensional cubic
lattice\cite{metzner_vollhardt}],
and they interact via a screened Coulomb interaction with the localized
electrons (that is described by an interaction strength $U$ between
electrons that are located at the same lattice site). All
energies are measured in units of $t^*$.  The Hamiltonian is
\begin{eqnarray}
H =-\frac{t^*}{2\sqrt{d}}\sum_{\langle i,j\rangle}d_i^{\dagger}d_j +&&E_f
\sum_i w_i-\mu\sum_i(d_i^{\dagger}d_i+w_i)\nonumber\\
&&+U\sum_id_i^{\dagger}d_iw_i,
\label{eq: ham}
\end{eqnarray}
where $d_i^{\dagger}$ $(d_i)$ is the spinless conduction electron creation
(annihilation) operator at lattice site $i$ and $w_i=0$ or 1 is a classical
variable corresponding to the localized $f$-electron number at site $i$. 
We will adjust
both $E_f$ and $\mu$ so that the average filling of the $d$-electrons is 1/2 
and the average filling of the $f$-electrons is 1/2 ($\mu=U/2$ and $E_f=0$).

In this contribution we will show how to derive the nonresonant, mixed, and
resonant Raman response of a system that crosses through a metal-insulator
transition by solving the problem exactly in the large dimensional limit
(employing dynamical mean-field theory).  In the case of nearest-neighbor 
hopping on a hypercubic lattice in infinite dimensions, we show
that the $A_{\rm 1g}$ response includes contributions from all 
processes, the $B_{\rm 1g}$ response is resonant or nonresonant,
and the $B_{\rm 2g}$ response is
purely resonant.  We provide computational results only for the nonresonant
$A_{\rm 1g}$ and $B_{\rm 1g}$ responses.  Resonant (and
mixed) Raman
scattering results will be presented elsewhere.  We also prove the 
Shastry-Shraiman relation, motivate the origin of the anomalous
features in the nonresonant $B_{\rm 1g}$ response, and show why they are 
not seen in the nonresonant $A_{\rm 1g}$ channel.

We present the formalism in Section II which includes a detailed analysis
of the Raman response in the different symmetry channels.  In Section III
we present the results of our calculations which display anomalous 
behavior for $B_{\rm 1g}$ and ordinary behavior for $A_{\rm 1g}$.  Our 
conclusions are presented in Section IV.

\section{Formalism}

\subsection{Nonresonant Raman response}

\begin{figure}
\vspace{5mm}
\epsfxsize=3.0in
\centerline{\epsffile{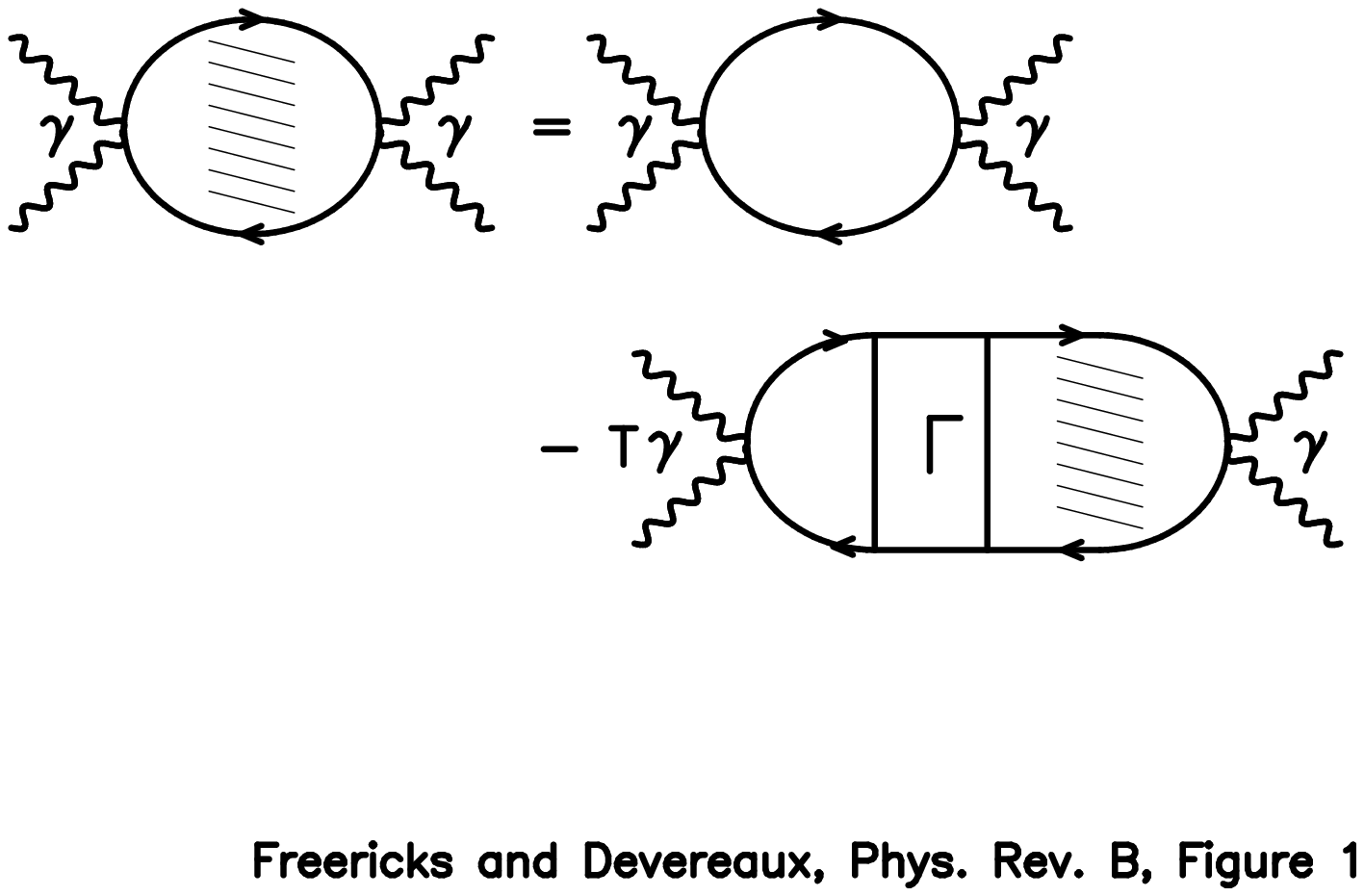}}
\vspace{.5cm}
\caption[]{Dyson equation for the nonresonant Raman response function.
Solid lines denote electron propagators and wavy lines denote
photon propagators. The shading denotes the fully renormalized susceptibility
and the symbol $\Gamma$ is the irreducible frequency-dependent charge vertex.
The vertex function $\gamma$ is the Raman scattering amplitude, which determines
the symmetry of the Raman scattering channel.
\label{fig: nonresonant}}

\end{figure}                                                                   

The Feynman diagrams for the nonresonant Raman response are shown in
Fig.~\ref{fig: nonresonant}.  The straight lines denote the momentum-dependent
Green's function and the wavy lines denote the photon propagator.  The 
shading is used to represent the renormalized susceptibility and the
symbol $\Gamma$ denotes the local irreducible charge vertex.  The functions
$\gamma({\bf k})$ are called the Raman scattering amplitudes, and they are 
chosen to have a well-defined spatial symmetry and no frequency dependence.
The Raman response is found from this frequency-dependent
density-density correlation function which is
\begin{eqnarray}
\chi_{\rm Raman}(i\nu_l)&=&\sum_{\bf k}\int_0^{\beta}d\tau e^{i\nu_l\tau}
\cr
&\Biggr\{&\frac{{\rm Tr}T_\tau\langle e^{-\beta H}\rho_{\bf k}(\tau)\rho_{\bf k}
(0)
\rangle}{Z}\cr
&-&\Biggr [ \frac{{\rm Tr}\langle e^{-\beta H}\rho_{\bf k}(0)\rangle}
{Z} \Biggr ]^2\Biggr\},
\label{eq: raman}
\end{eqnarray}
with the uniform (${\bf q}=0$) Raman density operator
\begin{equation}
\rho_{\bf k} = \gamma({\bf k})d^{\dagger}_{\bf k}d_{\bf k},
\quad d_{\bf k}=\frac{1}{N}\sum_j e^{-{\bf R}_j\cdot {\bf k}}d_j,
\label{eq: ramandensity}
\end{equation}
$Z=Tr\langle e^{-\beta H}\rangle$, the partition function,
and $i\nu_l=2i\pi lT$ the Bosonic Matsubara frequency (the $\tau$-dependence
of the operators is with respect to the full Hamiltonian). 
The Raman response is characterized in terms of the different spatial 
symmetries of the scattering amplitude.  One can expand
this function in a Fourier series and examine the contributions of the lowest
components of the series, and compare them to experiment. Alternatively, one can
start from a metallic Hamiltonian, and expand the Raman response in
powers of the electronic vector potential.  Under the assumption that the 
photon wavelength is much larger than the lattice spacing,
this latter approach yields
the following results for the Raman scattering amplitudes in two dimensions:
\begin{eqnarray}
\gamma_{A_{\rm 1g}}({\bf k})&\approx&\frac{\partial^2\epsilon({\bf k})}
{\partial k_x^2}+
\frac{\partial^2\epsilon({\bf k})} {\partial k_y^2}\approx
-\epsilon({\bf k}),\cr
\gamma_{B_{\rm 1g}}({\bf k})&\approx&\frac{\partial^2\epsilon({\bf k})}
{\partial k_x^2}-
\frac{\partial^2\epsilon({\bf k})} {\partial k_y^2}\approx
\cos k_x-\cos k_y,\cr
\gamma_{B_{\rm 2g}}({\bf k})&\approx&\frac{\partial^2\epsilon({\bf k})}
{\partial k_x\partial k_y}\approx
\sin k_x\sin k_y,
\label{eq: ramanvertex}
\end{eqnarray}
with $\epsilon({\bf k})\propto \cos k_x+\cos k_y$ the electronic 
bandstructure.  Note that the
B$_{\rm 2g}$ response vanishes for pure nearest-neighbor hopping, which is
what we consider here (the band structure is just a sum of cosines; hence all 
mixed derivatives vanish).  {\it The nonresonant $B_{\rm 2g}$ Raman response 
then vanishes, because the lowest-order Raman scattering amplitude vanishes.}

The above forms can be generalized to the infinite-d limit for non-resonant
scattering by noting
that $\gamma({\bf k})$ satisfies\cite{abrikosov} 
\begin{equation}
\gamma({\bf k})=\sum_{\alpha\beta}e_{i\alpha}\frac{\partial^2\epsilon({\bf k})}
{\partial k_\alpha\partial k_\beta}e_{o\beta},
\label{eq: gamma_def}
\end{equation}
with ${\bf e}_i$ (${\bf e}_o$) the incoming (outgoing) photon polarization.
In systems with hopping beyond nearest neighbors, it is not possible to
project completely onto the $A_{\rm 1g}$ sector with just one choice of
polarizations, since either $B_{\rm 1g}$ or $B_{\rm 2g}$ sectors will always
be mixed in; it is possible to project onto $B_{\rm 1g}$ or $B_{\rm 2g}$
with one measurement.  In the infinite-dimensional limit, with nearest-neighbor
hopping only, we can choose 
${\bf e}_{i\alpha}={\bf e}_{o\alpha}=1$ for $A_{\rm 1g}$,
${\bf e}_{i\alpha}=1$, ${\bf e}_{o\alpha}=(-1)^\alpha$ for $B_{\rm 1g}$, and
${\bf e}_{i\alpha={\rm even}}=0$, ${\bf e}_{i\alpha={\rm odd}}=1$,
${\bf e}_{o\alpha={\rm even}}=1$, ${\bf e}_{o\alpha={\rm odd}}=0$ for
$B_{\rm 2g}$.  The resulting Raman scattering amplitudes are
\begin{equation}
\gamma_{A_{\rm 1g}}({\bf k})\approx c-\epsilon({\bf k}),\quad
\gamma_{B_{\rm 1g}}({\bf k})\approx
\frac{1}{\sqrt{d}}\sum_{i=1}^d(-1)^i\cos k_i,
\label{eq: ramanvertex_infd}
\end{equation}
and $\gamma_{B_{\rm 2g}}({\bf k})=0$.  Note that
we include a constant term $c$ in the A$_{\rm 1g}$ amplitude, since it is
allowed by symmetry.  

The nonresonant Raman response in the B$_{\rm 1g}$ channel has
no vertex corrections\cite{khurana,raman_us} 
and is equal to the bare bubble. The key element needed to show this
is that the irreducible charge vertex $\Gamma$ is independent of 
momentum (since it is local).
Hence, we must evaluate a summation over {\bf k} of the form
\begin{equation}
\sum_{\bf k} \frac{1}{\sqrt{d}}\sum_{i=1}^d(-1)^i\cos k_i \frac{1}{X+
\frac{2}{\sqrt{d}} \sum_{j=1}^d\cos k_j},
\label{eq: b1gsum}
\end{equation}
which arises when solving the Dyson equation for the B$_{\rm 1g}$ response
(and we can assume without loss of generality that
the imaginary part of $X$ is greater than zero).
Writing the fraction in Eq.~(\ref{eq: b1gsum}) as the integral of an
exponential\cite{mueller-hartmann}
\begin{eqnarray}
&&\frac{1}{X+\frac{2}{\sqrt{d}} \sum_{j=1}^d\cos k_j}
=\cr
&&-i\int_0^{\infty}dz\exp[iz(X+\frac{2}{\sqrt{d}} \sum_{j=1}^d\cos k_j)],
\label{eq: expint}
\end{eqnarray}
allows us to decouple the summation over momentum to the sum over $d$
identical terms, each multiplied by $(-1)^i$.  This then vanishes for all even
$d$ and for odd-$d$
in the limit $d\rightarrow\infty$ (due to the $1/\sqrt{d}$ term).       So
the nonresonant $B_{\rm 1g}$ response reduces to the evaluation of the
bare bubble.

In the $A_{\rm 1g}$ channel, the Raman scattering amplitude has the 
same symmetry as the irreducible charge vertex, so the corresponding
summation over momentum does not vanish, and the nonresonant $A_{\rm 1g}$
Raman response is renormalized by the irreducible charge vertex.

The Falicov-Kimball model can be solved exactly in the infinite-dimensional
limit by using dynamical mean-field theory (see Ref. \onlinecite{bf} for details). 
We summarize
the main points to establish our notation.  The local Green's function at
the Fermionic Matsubara frequency $i\omega_n=i\pi T(2n+1)$ is defined by
\begin{eqnarray}
G_n&=&G(i\omega_n)\cr
&=&-{\rm Tr}T_{\tau}\int_{0}^{\beta}d\tau
e^{i\omega_n \tau}\frac{\langle
e^{-\beta H_{at}} d(\tau)d^{\dagger}(0)S(\lambda) \rangle}{Z},
\label{eq: gdef}
\end{eqnarray}
with
\begin{equation}
Z=Z_0(\mu)+e^{-\beta(E_f-\mu)}Z_0(\mu-U),
\label{eq: zdef}
\end{equation}
the atomic partition function expressed in terms of
\begin{equation}
Z_0(\mu)={\rm Tr_d}\langle e^{-\beta H_0}S(\lambda)\rangle,\quad
H_0=-\mu d^{\dagger}d.
\label{eq: z0def1}
\end{equation}
In the above equations, the atomic Hamiltonian
$H_{at}$ is the Hamiltonian of Eq.~(\ref{eq: ham})
restricted to one site, with $t^*=0$, and all time dependence is with respect
to $H_{at}$.  The evolution operator $S(\lambda)$ satisfies
\begin{equation}
S(\lambda)=\exp[-\int_0^{\beta}d\tau\int_0^{\beta}
d\tau^{\prime} d^{\dagger}(\tau)\lambda(\tau-\tau^{\prime})d(\tau^{\prime})],
\label{eq: sdef}
\end{equation}
with $\lambda(\tau-\tau^{\prime})$
a time-dependent atomic field adjusted to make
the atomic Green's function equal to the local lattice Green's function.  We
define an effective medium by
\begin{equation}
G_0^{-1}(i\omega_n)=G_n^{-1}+\Sigma_n=i\omega_n+\mu-\lambda_n,
\label{eq: g0def}
\end{equation}
with $\Sigma_n$ the local self-energy and $\lambda_n$ the Fourier transformation
of $\lambda(\tau)$.  The trace in Eq.~(\ref{eq: gdef}) can be evaluated
directly to yield
\begin{equation}
G_n=w_0G_0(i\omega_n)+w_1[G_0^{-1}(i\omega_n)-U]^{-1},
\label{eq: gatomic}
\end{equation}
with $w_0=1-w_1$,
\begin{equation}
w_1=\exp[-\beta(E_f-\mu)]Z_0(\mu-U)/Z,
\label{eq: w1def}
\end{equation}
and 
\begin{equation}
Z_0(\mu)=2e^{\beta\mu/2}\prod_{n=-\infty}^{\infty}
\frac{i\omega_n+\mu-\lambda_n}{i\omega_n}.
\label{eq: z0def}
\end{equation}
The self-consistency relation needed to determine $\lambda_n$ and $G_n$
is to equate the local lattice Green's function to the atomic Green's function
via
\begin{equation}
G_n=\int_{-\infty}^{\infty}d\epsilon\frac{\rho(\epsilon)}
{i\omega_n+\mu-\Sigma_n-\epsilon},
\label{eq: glat}
\end{equation}
with $\rho(\epsilon)=\exp(-\epsilon^2)/\sqrt{\pi}$ the noninteracting density
of states for the infinite-dimensional hypercubic lattice.

The iterative algorithm to solve for $G_n$ starts with $\Sigma_n=0$.  Then
Eq.~(\ref{eq: glat}) is used to find $G_n$, Eq.~(\ref{eq: g0def}) is employed
to extract the effective medium, Eq.~(\ref{eq: gatomic}) is used to find a
new local Green's function, and then Eq.~(\ref{eq: g0def}) is used to find
the new self-energy. This sequence of steps is then repeated until it converges,
which usually requires only about a dozen or so iterations. This algorithm
can also be used on the real axis (with suitably modified equations) to
directly solve for the Green's function and self-energy on the real axis.

Once the Green's functions are known, then the nonresonant Raman response
can be calculated directly.  The $B_{\rm 1g}$ channel is simple, since we
need only evaluate the bare bubble.  Noting that the average of the square
of the $B_{\rm 1g}$ Raman scattering amplitude is $1/2$ then yields
\begin{equation}
\chi_{B_{\rm 1g}}(i\nu_l)=-\frac{T}{2}\sum_n\frac{G(i\omega_n)-G(i\omega_{n+l})}
{i\nu_l+\Sigma(i\omega_n)-\Sigma(i\omega_{n+l})},
\label{eq: b1g}
\end{equation}
for the Raman response on the imaginary axis.
This formula can be easily analytically continued to the real axis by
following the same procedure outlined in the calculation
of the dynamical charge susceptibility~\cite{fk_charge}: rewrite the sum over
Matsubara frequencies by a contour integral of advanced or retarded Green's
functions and self-energies multiplied by the Fermi factor, and then deform the
contours to the real axis picking up any poles in the complex plane.  Under
the assumption that there are no extra poles when the contours are deformed,
one ends up with the following expression for the B$_{1g}$ response:
\begin{eqnarray}
\chi_{B_{\rm 1g}}(\nu)&=&\frac{-i}{4\pi}\int_{-\infty}^{\infty}d\omega\cr
&\Biggr\{&
f(\omega)\frac{G(\omega)-G(\omega+\nu)}{\nu+
\Sigma(\omega)-\Sigma(\omega+\nu)} \cr
&-&f(\omega+\nu)\frac{G^*(\omega)-G^*(\omega+\nu)}{\nu+
\Sigma^*(\omega)-\Sigma^*(\omega+\nu)} \cr
&-&[f(\omega)-f(\omega+\nu)]
\frac{G^*(\omega)-G(\omega+\nu)}{\nu+
\Sigma^*(\omega)-\Sigma(\omega+\nu)}\Biggr\},
\label{eq: b1greal}                
\end{eqnarray}
with $f(\omega)=1/[1+\exp(\beta\omega)]$ the Fermi function.
We verify that this expression is indeed accurate, by using the spectral
formula to calculate the Raman response on the imaginary axis and comparing it
to the result directly calculated from the expression in Eq.~(\ref{eq: b1g}).
We find that the results rarely differ by more than one part in a thousand
confirming the accuracy of the analytic continuation (we believe, but
cannot prove, that no additional poles exist in the complex plane, which
would mean the analytic continuation is exact).   

We can use the expression in Eq.~(\ref{eq: b1greal}) to prove the 
Shastry-Shraiman relation.  Using the identity that relates the local
Green's function on the lattice to the self energy in Eq.~(\ref{eq: glat})
(but evaluated on the real axis rather than the imaginary axis) yields
\begin{eqnarray}
&&\frac{G(\omega)-G(\omega+\nu)}{\nu+\Sigma(\omega)-\Sigma(\omega+\nu)}
=\cr
&&\int_{-\infty}^{\infty}d\epsilon \frac{\rho(\epsilon)}
{[\omega+\mu-\Sigma(\omega)-\epsilon][\omega+\nu+\mu-\Sigma(\omega+\nu)
-\epsilon]}.
\label{eq: b1g_identity}
\end{eqnarray}
Employing this identity, and related ones for the advanced Green's functions,
in Eq.~(\ref{eq: b1greal}) produces
\begin{eqnarray}
&&\chi_{B_{\rm 1g}}(\nu)=\frac{1}{2\pi}\int_{-\infty}^{\infty}d\omega
\int_{-\infty}^{\infty}d\epsilon\rho(\epsilon)
\cr
&&\Biggr\{\frac{f(\omega)}{\omega+\nu+\mu-\Sigma(\omega+\nu)-\epsilon}
{\rm Im}\frac{1}{\omega+\mu-\Sigma(\omega)-\epsilon}\cr
&&+\frac{f(\omega+\nu)}{\omega+\mu-\Sigma^*(\omega)-\epsilon}
{\rm Im}\frac{1}{\omega+\nu+\mu-\Sigma(\omega+\nu)-\epsilon}
\Biggr\}.
\label{eq: b1greal_2}
\end{eqnarray}
Using the definition of the spectral function 
\begin{equation}
A(\epsilon,\omega)=-\frac{1}{\pi}{\rm Im}\frac{1}
{\omega+\mu-\Sigma(\omega)-\epsilon},
\label{eq: spectraldef}
\end{equation}
and taking the imaginary part of Eq.~(\ref{eq: b1greal_2}), then yields
\begin{eqnarray}
{\rm Im}\chi_{B_{\rm 1g}}(\nu)&=&2\pi\int_{-\infty}^{\infty}d\omega
\int_{-\infty}^{\infty}d\epsilon\rho(\epsilon)
\cr
&[&f(\omega)-f(\omega+\nu)]A(\epsilon,\omega)A(\epsilon,\omega+\nu),\cr
&\propto&\nu\sigma(\nu),
\label{eq: shastry-shraiman}
\end{eqnarray}
which is the Shastry-Shraiman relation---the imaginary part of the nonresonant
$B_{\rm 1g}$
Raman response is proportional to the frequency $\nu$ times the optical 
conductivity
$\sigma(\nu)$.  This conclusion comes from comparing the integration in 
Eq.~(\ref{eq: shastry-shraiman}) to the well-known result for the 
optical conductivity on the infinite-d hypercubic lattice\cite{jarrell_oc}.
Note that this final formula for the Raman response depends only on the
shape of the spectral function.  Since the derivation was model-independent
(like the derivation of the optical conductivity\cite{jarrell_oc}), this form 
for the nonresonant $B_{\rm 1g}$ Raman response holds for all models. 
(Similarly, in models with next-nearest-neighbor hopping, one can also
show that the nonresonant B$_{\rm 2g}$ response
also satisfies the Shastry-Shraiman
relation.)  We note
that this is a consequence of the momentum-independent self energy---any
dependence of the self energy on momentum will generally violate the 
Shastry-Shraiman relation, although it may still be an accurate approximation.

\begin{figure}[htbp]
\epsfxsize=3.0in
\centerline{\epsffile{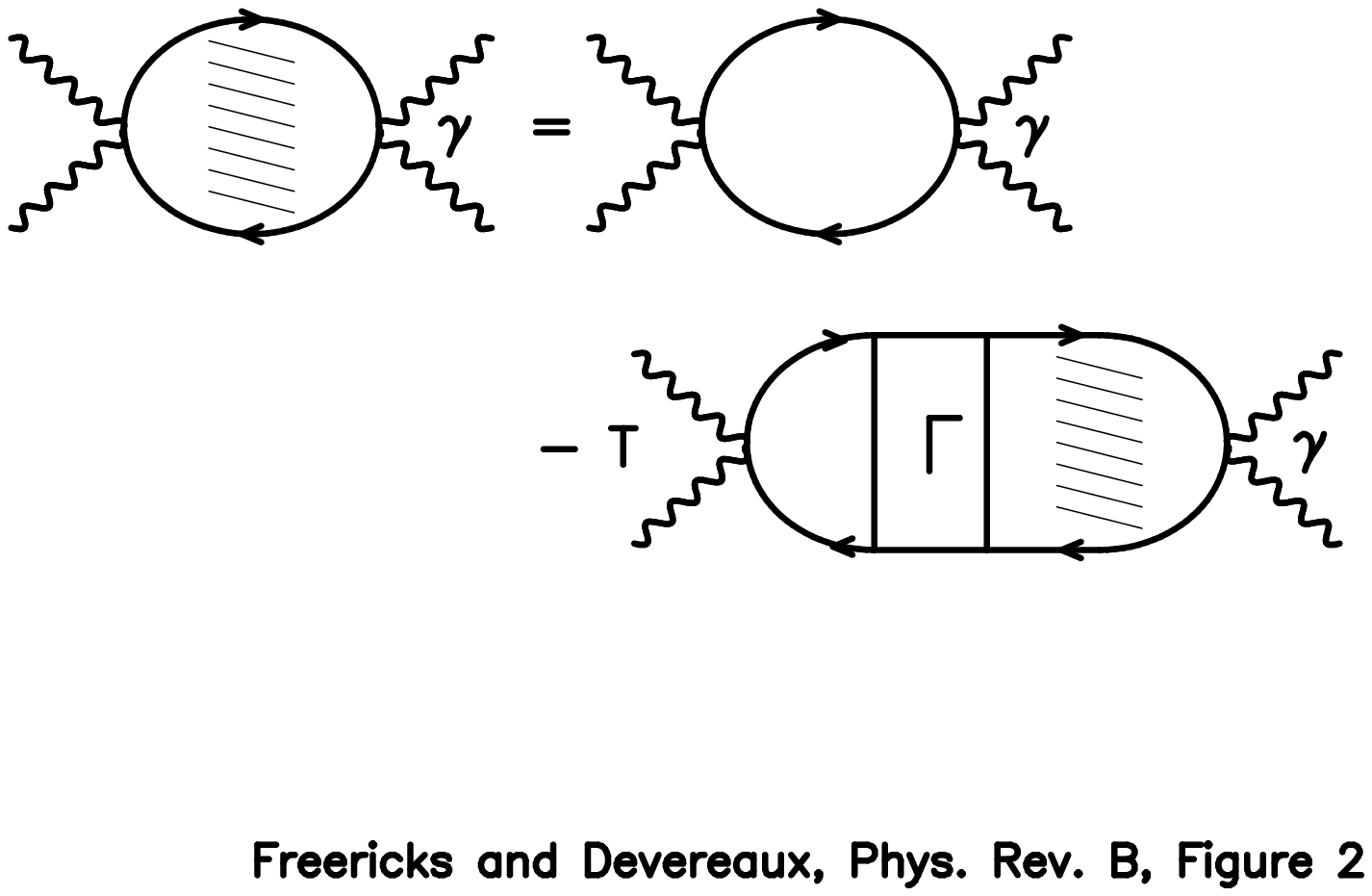}}
\caption{Supplemental
Feynman diagrams for the nonresonant $A_{\rm 1g}$ Raman response.
These diagrams are identical to those in Figure 1 except they have one
fewer power of $\gamma$.
\label{fig: a1g_feynman}}
\end{figure}

The A$_{\rm 1g}$ response is more complicated, because it requires a proper
treatment of the vertex contributions.  Fortunately, the charge vertex
for the Falicov-Kimball model is well-known \cite{fk_charge} and assumes
a simple form (for $i\nu_l\ne 0$)
\begin{equation}
\Gamma(i\omega_m,i\omega_n;i\nu_l)=\delta_{m,n}\frac{1}{T}\frac{
\Sigma(i\omega_n)-\Sigma(i\omega_{n+l})}{G(i\omega_n)-G(i\omega_{n+l})}.
\label{eq: chargevertex}
\end{equation}
Hence, the Raman response in the A$_{\rm 1g}$ channel can be found by solving 
the relevant Dyson's equation, using the above form of the charge vertex.  
The set of Feynman diagrams is shown in Figs.~\ref{fig: nonresonant} and
\ref{fig: a1g_feynman}. Note
that we have to solve these two coupled equations for the $A_{\rm 1g}$
Raman response.

The difference between the two diagrams shown in Figs.~\ref{fig: nonresonant}
and \ref{fig: a1g_feynman}
is the number of
factors of the Raman scattering amplitude $\gamma$; the nonresonant
Raman response requires the dressed response that has only one power of 
$\gamma$---this dressed response satisfies a simple Dyson equation, shown
in Fig.~\ref{fig: a1g_feynman}.  We denote the dressed response function
in Fig.~\ref{fig: a1g_feynman} by 
$\chi^{\prime}(i\omega_m,i\omega_n;i\nu_l)$.
Then a straightforward evaluation of the Feynman diagrams and a solution
of the Dyson equation produces
\begin{eqnarray}
\chi^{\prime}(i\omega_m,i\omega_n;i\nu_l)&&=\cr
&&\frac{\delta_{mn}\chi_0^{\prime}(i\omega_m;i\nu_l)}
{1+\chi_0(i\omega_m;i\nu_l)T\Gamma(i\omega_m,i\omega_m;i\nu_l)},
\label{eq: a1g_chiprime}
\end{eqnarray}
with  the irreducible charge vertex function found in 
Eq.~(\ref{eq: chargevertex}),
\begin{equation}
\chi_0(i\omega_n;i\nu_l)=-\frac{G(i\omega_n)-G(i\omega_{n+l})}
{i\nu_l+\Sigma(i\omega_n)-\Sigma(i\omega_{n+l})},
\label{eq: chi0def}
\end{equation}
and
\begin{equation}
\chi_0^{\prime}(i\omega_n;i\nu_l)=-\frac{c(G_n-G_{n+l})-Z_nG_n+Z_{n+l}G_{n+l}}
{i\nu_l+\Sigma_n-\Sigma_{n+l}}.
\label{eq: chi0primedef}
\end{equation}
Here we used the notation $Z_n=i\omega_n+\mu-\Sigma(i\omega_n)$. Now, knowledge 
of $\chi^{\prime}$ allows us to solve for the Raman response in 
Fig.~\ref{fig: nonresonant}.
The end result is
\begin{eqnarray}
&&\chi_{A_{\rm 1g}}(i\nu_l)=
T\sum_n\cr
&&\frac{\bar\chi_0(i\omega_n;i\nu_l)-
[\chi_0(i\omega_n;i\nu_l)+G_nG_{n+l}] T\Gamma(i\omega_n,i\omega_n;i\nu_l)}
{1+\chi_0(i\omega_n;i\nu_l)T\Gamma(i\omega_n,i\omega_n;i\nu_l)},
\label{eq: a1g}
\end{eqnarray}
where the charge vertex is found in Eq.~(\ref{eq: chargevertex}), the bare
susceptibility $\chi_0$  is found in Eq.~(\ref{eq: chi0def})
and the other bare susceptibility $\bar\chi_0$ (which is where all of the $c$
dependence lies) is
\begin{eqnarray}
\bar\chi_0(i\omega_n;i\nu_l)&=&[-c^2(G_n-G_{n+l})\cr
&&+2c(Z_nG_n-Z_{n+l}G_{n+l})\cr
&+&Z_n-Z_{n+l}
-Z_n^2G_n+Z_{n+l}^2G_{n+l}]\cr
&&/[i\nu_l+\Sigma_n-\Sigma_{n+l}].
\label{eq: chi0bardef}
\end{eqnarray}
Surprisingly, all of the $c$ dependence actually drops out of the
nonresonant
$A_{\rm 1g}$ Raman response (as it does in conventional
metals\cite{metals}).  This can be seen by substituting the 
representations for $\chi_0$ and $\Gamma$ into the denominator
in the first term of Eq.~(\ref{eq: a1g}).  One finds that the summation
for the first term
can then be written in the general form $\sum_n(Y_n-Y_{n+l})/i\nu_l$, which
vanishes since the individual summations converge and one can change the
summation index $n\rightarrow n+l$ in the first term [we similarly note that
$\sum_n(\Sigma_n-\Sigma_{n+l})=0$]. Hence we arrive at
the final form for the nonresonant $A_{\rm 1g}$ Raman response:
\begin{eqnarray}
\chi_{A_{\rm 1g}}(i\nu_l)&=&\frac{T}{i\nu_l}\sum_n
\frac{(\Sigma_n-\Sigma_{n+l})G_nG_{n+l}}{\chi_0(i\omega_n;i\nu_l)}
\cr
&=&\frac{T}{i\nu_l}\sum_n\frac{(\Sigma_n-\Sigma_{n+l}) 
(i\nu_l+\Sigma_n-\Sigma_{n+l})}{G_{n+l}^{-1}-G_n^{-1}} .
\label{eq: a1g_final}
\end{eqnarray}
It is a straightforward exercise to perform a similar analytic continuation of
this expression. The result is
\begin{eqnarray}
&&\chi_{A_{\rm 1g}}(\nu)=\frac{i}{2\pi\nu}\int_{-\infty}^{\infty}d\omega\cr
&&\Biggr\{
f(\omega)\frac{[\Sigma(\omega)-\Sigma(\omega+\nu)][\nu+\Sigma(\omega)-
\Sigma(\omega+\nu)]}{G^{-1}(\omega+\nu)-G^{-1}(\omega)} \cr
&&-f(\omega+\nu)\frac{[\Sigma^*(\omega)-\Sigma^*(\omega+\nu)][\nu+
\Sigma^*(\omega)-\Sigma^*(\omega+\nu)]}{G^{-1*}(\omega+\nu)-G^{-1*}(\omega)}
\cr
&&-[f(\omega)-f(\omega+\nu)]\cr
&&\times
\frac{[\Sigma^*(\omega)-\Sigma(\omega+\nu)][\nu+\Sigma^*(\omega)-
\Sigma(\omega+\nu)]}{G^{-1}(\omega+\nu)-G^{-1*}(\omega)}
\Biggr\}.
\label{eq: a1greal}
\end{eqnarray} 

\subsection{Resonant Raman response}

\begin{figure}
\epsfxsize=3.0in
\centerline{\epsffile{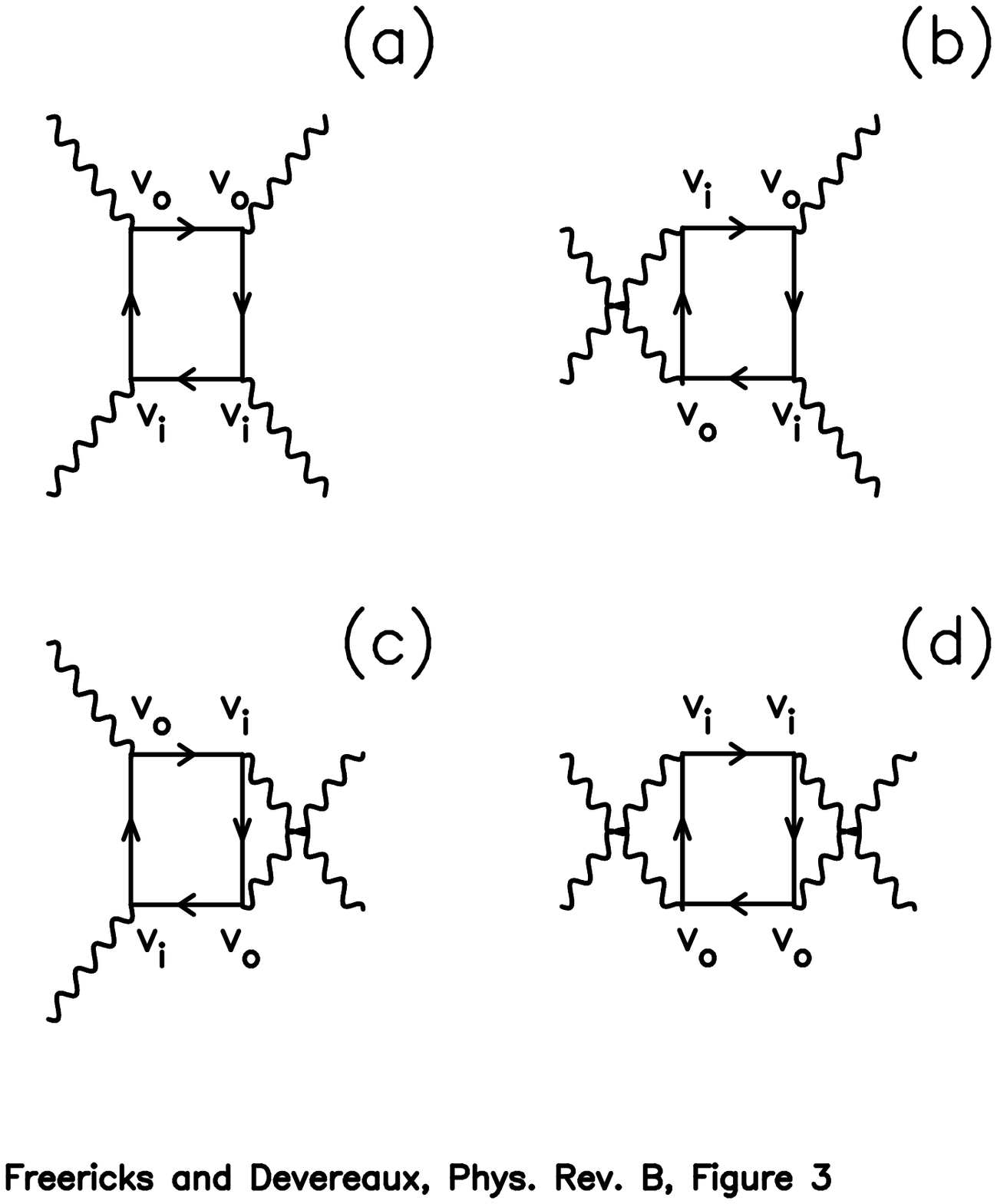}}
\caption{Bare resonant Raman scattering diagrams.  The (a) uncrossed,
(b) first partially crossed, (c) second partially crossed, and (d)
fully crossed diagrams are all shown.  The straight lines are
momentum dependent Green's functions, the wiggly lines are photon 
propagators, and the vertex factors are the dot product of the 
photon polarization with the Fermi velocity as described in the text
($v_i={\bf e}_i\cdot {\bf v}$, $v_o={\bf e}_o\cdot {\bf v}$).
Note that the two partially crossed diagrams (b) and (c) are equal to each
other.
\label{fig: resonant}}
\end{figure}

Resonant Raman scattering arises from a fourth-order process with respect to the
photon vector potential.  One way to view this scattering is to ``separate'' the
two-photon-two-electron vertex of nonresonant Raman scattering in
Fig.~\ref{fig: nonresonant} into two single-photon-two-electron vertices.
Keeping in mind the direct and exchange possibilities, results in four
distinct bare resonant Raman scattering diagrams, which are depicted in
Fig.~\ref{fig: resonant}: (a) the uncrossed (direct) diagram; (b) the first
partially-crossed diagram; (c) the second partially crossed diagram; and (d)
the fully crossed (exchange) diagram.  In all relevant cases, the partially
crossed diagrams (b) and (c) are equal to each other.

In general, the Raman vertices for resonant scattering involve expectation 
values of the momentum operator between the conduction band and the excited
states.  However, in the single-band model, 
the Raman vertices in Fig.~\ref{fig: resonant} are equal to the dot product of
the photon polarization vector {\bf e} with the Fermi velocity {\bf v}
\begin{equation}
v_\alpha=\frac{\partial \epsilon({\bf k})}{\partial k_\alpha}=2\sin k_\alpha.
\label{eq: fermi_velocity}
\end{equation}
Again, 
we distinguish between the incoming photon polarization ${\bf e}_i$ and the
outgoing photon polarization ${\bf e}_o$.  The different symmetry channels
are projected by appropriate choice of the polarization vectors.  
We  choose ${\bf e}_{i\alpha}={\bf e}_{o\alpha}=1$ for $A_{\rm 1g}$,
${\bf e}_{i\alpha}=1$, ${\bf e}_{o\alpha}=(-1)^\alpha$ for $B_{\rm 1g}$, and
${\bf e}_{i\alpha={\rm even}}=0$, ${\bf e}_{i\alpha={\rm odd}}=1$,
${\bf e}_{o\alpha={\rm even}}=1$, ${\bf e}_{o\alpha={\rm odd}}=0$ for
$B_{\rm 2g}$. 
It is important to note
that in all cases, a single factor of ${\bf e}\cdot {\bf v}$ is an odd function
of {\bf k}, so the only way to get nonzero summations over momentum is to have
an even number of ${\bf e}\cdot {\bf v}$ factors in any given integration.

\begin{figure}
\epsfxsize=3.0in
\centerline{\epsffile{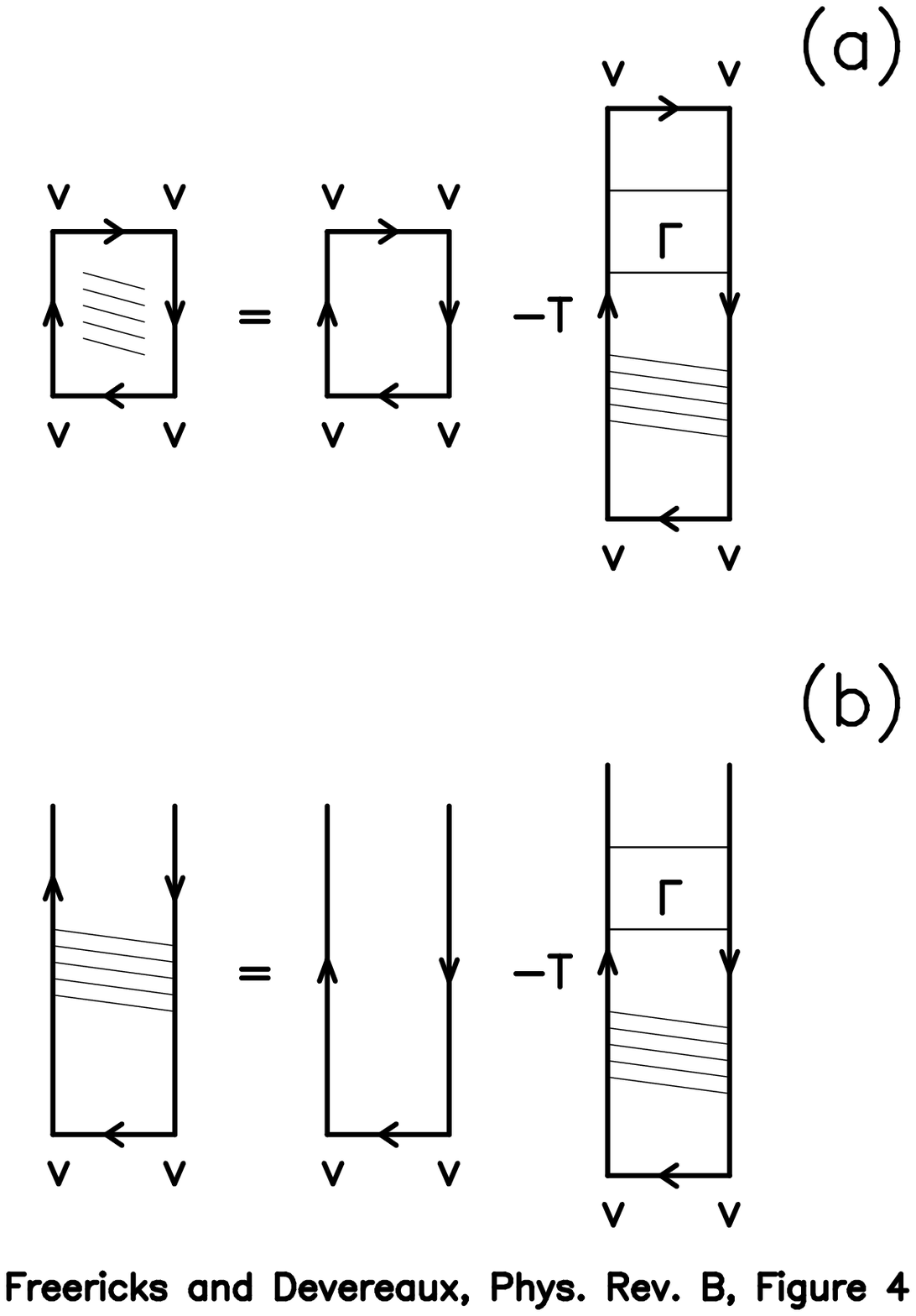}}
\caption{Renormalized resonant Raman scattering diagrams for Figures 3(a)
and 3(d) in the $B_{\rm 1g}$
and $B_{\rm 2g}$ channels.  We suppress the photon lines that are
explicitly shown in Fig.~\ref{fig: resonant}, for simplicity.  In (a)
one has the diagrams for the Raman response, while in (b) the supplemental
Dyson equation needed to solve for the dressed Raman response is given.
The only difference for the $B_{\rm 1g}$ and $B_{\rm 2g}$ channels is
the choice of polarization vectors ${\bf e}_i$ and ${\bf e}_o$, and we have
suppressed the incoming and outgoing indices on the vertices $v$. The symbol
$\Gamma$ denotes the local irreducible charge vertex. Note that only the
direct and exchange diagrams [Figs.~\ref{fig: resonant}(a) and (d)] are
renormalized.
\label{fig: renorm_resonant}}
\end{figure}

In general, the bare resonant Raman scattering diagrams must be renormalized
by attaching appropriate irreducible charge vertex functions (and higher-order
generalizations, if possible) to all relevant Green's function legs.  Since all
${\bf e}\cdot {\bf v}$ factors are odd in {\bf k} and the irreducible charge
vertex has $A_{\rm 1g}$ symmetry, there can be no renormalization of any 
single vertex.  Similarly, one can argue that there are no three-particle
or four-particle vertex renormalizations possible either.  The only possibility
is a two-particle vertex that connects opposite Green's function lines.
In the $A_{\rm 1g}$ resonant Raman scattering channel, all possible vertical
and horizontal renormalizations of each bare diagram is possible.  For the 
$B_{\rm 1g}$ and $B_{\rm 2g}$ channels, there are simplifications.  One can
quickly verify that in both of those cases the product ${\bf e}_i\cdot {\bf v}
{\bf e}_o\cdot {\bf v}$ is orthogonal to the $A_{\rm 1g}$ symmetry, so the
partially crossed diagrams (b) and (c) cannot have any renormalization
either.  Furthermore, this also implies that attaching a vertical charge vertex
to either the (a) or the (d) diagram vanishes for the same reason.  But
a horizontal attachment of the vertex is possible in both the (a) and (d)
diagrams.  Hence,
the renormalized resonant Raman scattering in both the $B_{\rm 1g}$ and
$B_{\rm 2g}$ channels is described by the coupled set of equations in 
Fig.~\ref{fig: renorm_resonant}.  A similar, but more complicated, set of
diagrams is needed for the $A_{\rm 1g}$ channel, where renormalizations are
present on both the vertical and horizontal pairs of legs for all diagrams.  
We don't show those Feynman diagrams here.

Evaluation of these diagrams, and their analytic continuation to the real
axis is tedious.  We leave that task for a future publication.  But we do
note that both of the renormalized diagrams (a) and (d) of 
Fig.~\ref{fig: resonant} will have the zero-frequency ($i\nu_l=0$) piece
of the irreducible charge vertex renormalizing them.  This is the nondiagonal
piece of the charge vertex, and it can get large when the system is
tuned to lie near a phase-separation transition.  Hence, we expect the
resonant Raman scattering to be enhanced whenever one is close to a
phase separation. Note that this enhancement will occur for all photon
frequencies, since the zero-frequency vertex couples all such diagrams; this
implies that one will not see this effect by tuning through a resonant
frequency.

\subsection{Mixed Raman response}

\begin{figure}
\epsfxsize=3.0in
\centerline{\epsffile{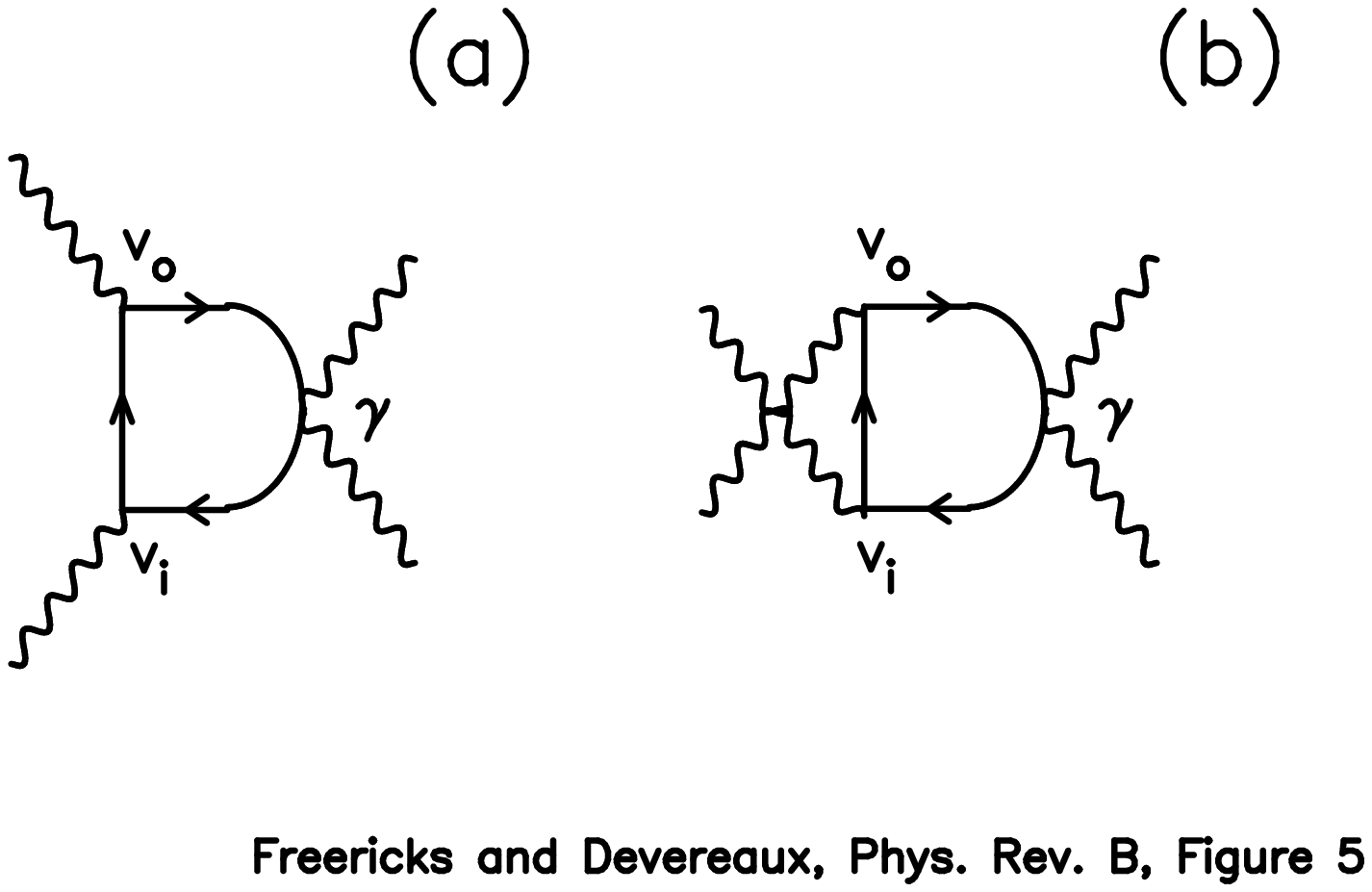}}
\caption{Bare mixed Raman response.  The single-photon vertices are multiplied
by the respective factor of the polarization vector dotted into the 
Fermi velocity, while the two-photon vertex is multiplied by the corresponding
Raman scattering amplitude.  Note that no renormalization is possible for the
$B_{\rm 1g}$ channel, but the diagram is renormalized in the $A_{\rm 1g}$
channel.
\label{fig: mixed}}
\end{figure}

The mixed Raman response comes from the cross terms between the linear and
quadratic terms in the vector potential.  As such, these diagrams have two
single-photon-two-electron vertices and one two-photon-two-electron vertex.
Since the $B_{\rm 2g}$
Raman scattering amplitude vanishes for nearest-neighbor hopping
on a hypercubic lattice, there is no mixed Raman response for that
channel.  The bare mixed Raman response is shown in Fig.~\ref{fig: mixed}.
There are two possible diagrams corresponding to the direct or exchange
processes.  In the $B_{\rm 1g}$ channel, the dressed mixed response is equal to 
the bare mixed
response, since the irreducible charge vertex cannot be inserted anywhere.
For the $A_{\rm 1g}$ channel, the bare mixed response is renormalized by the
irreducible charge vertex in a similar way to how it renormalized the
nonresonant Raman response.  We don't present any numerical results for the
mixed Raman response here, but will do so in another publication.  It turns out,
that one can show that in the $B_{\rm 1g}$ channel with nearest-neighbor
hopping only, the mixed response is
a $1/d$ correction and can be neglected; it cannot be neglected for the
$A_{\rm 1g}$ sector. The $1/d$ correction for $B_{\rm 1g}$ arises from the
fact that the summation over momentum has a ``form factor'' proportional to
$\cos(k_i)(-1)^i\sin^2(k_j)(-1)^j$.  This term cancels when summed over $i$ and
$j$ except for $i=j$.  This latter constraint forces the mixed diagram to
be a $1/d$ correction.  In the $A_{\rm 1g}$ case, there are no factors of $(-1)$
so the terms with all $i$ and $j$ contribute, and it is an $O(1)$ term.

\section{Results}

\begin{figure}
\vspace{5mm}
\epsfxsize=3.0in
\centerline{\epsffile{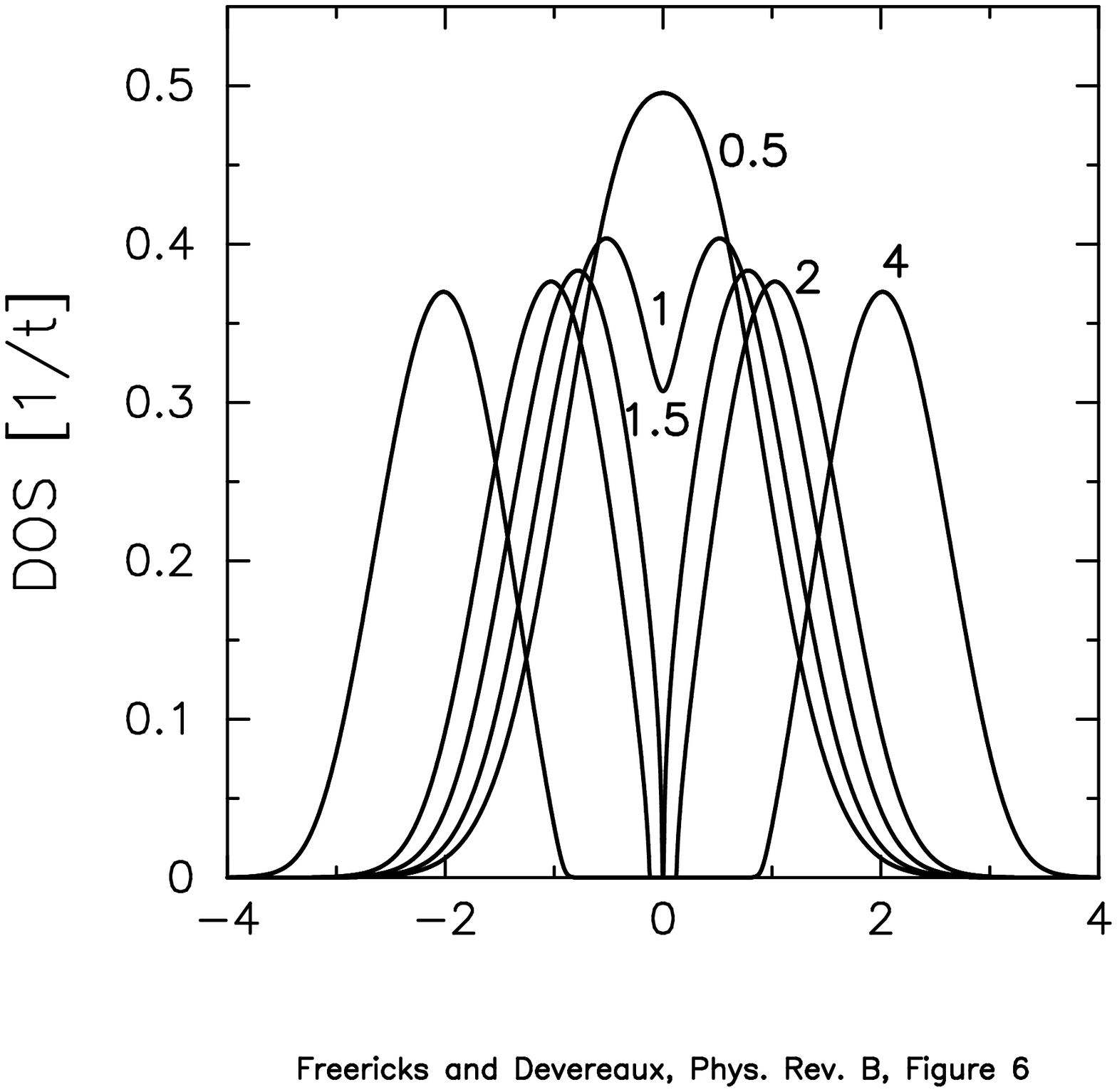}}
\vspace{.5cm}
\caption[]{Interacting density of states for the Falicov-Kimball model.
Results are shown for $U=$ 0.5, 1, 1.5, 2, and 4 (the numbers in the figure
label the value of $U$).  Note how the system
first develops a pseudogap (1.0) before the metal-insulator transition at
$U=1.5$.  The density of states is independent of temperature.
}
\label{fig: dos}
\end{figure}
 
The Falicov-Kimball model has a ground state that is not a Fermi liquid because
the lifetime of a quasiparticle is finite at the Fermi energy.  In addition,
the imaginary part of the self energy has the wrong sign of curvature
to be a Fermi liquid. We study the model at half filling. As $U$ 
increases, the system first enters a pseudogap phase, where spectral weight
is depleted near the chemical potential, and then undergoes a metal-insulator
transition (the pseudogap phase is possible because the ground state is
not a Fermi liquid).  The interacting density of states (DOS)
is, however, temperature-independent for fixed $U$ and fixed electron 
fillings\cite{vandongen_leinung}. It is plotted in Fig.~\ref{fig: dos} for a
range of values of $U$:
$U<0.65$ corresponds to a weakly-correlated
metal, while a pseudogap phase appears for $0.65<U<1.5$ moving through
a quantum critical point at $U=1.5$ to the insulator phase $U>1.5$ (we neglect
all possible charge-density-wave phases here). 

\begin{figure}
\vspace{5mm}
\epsfxsize=3.0in
\centerline{\epsffile{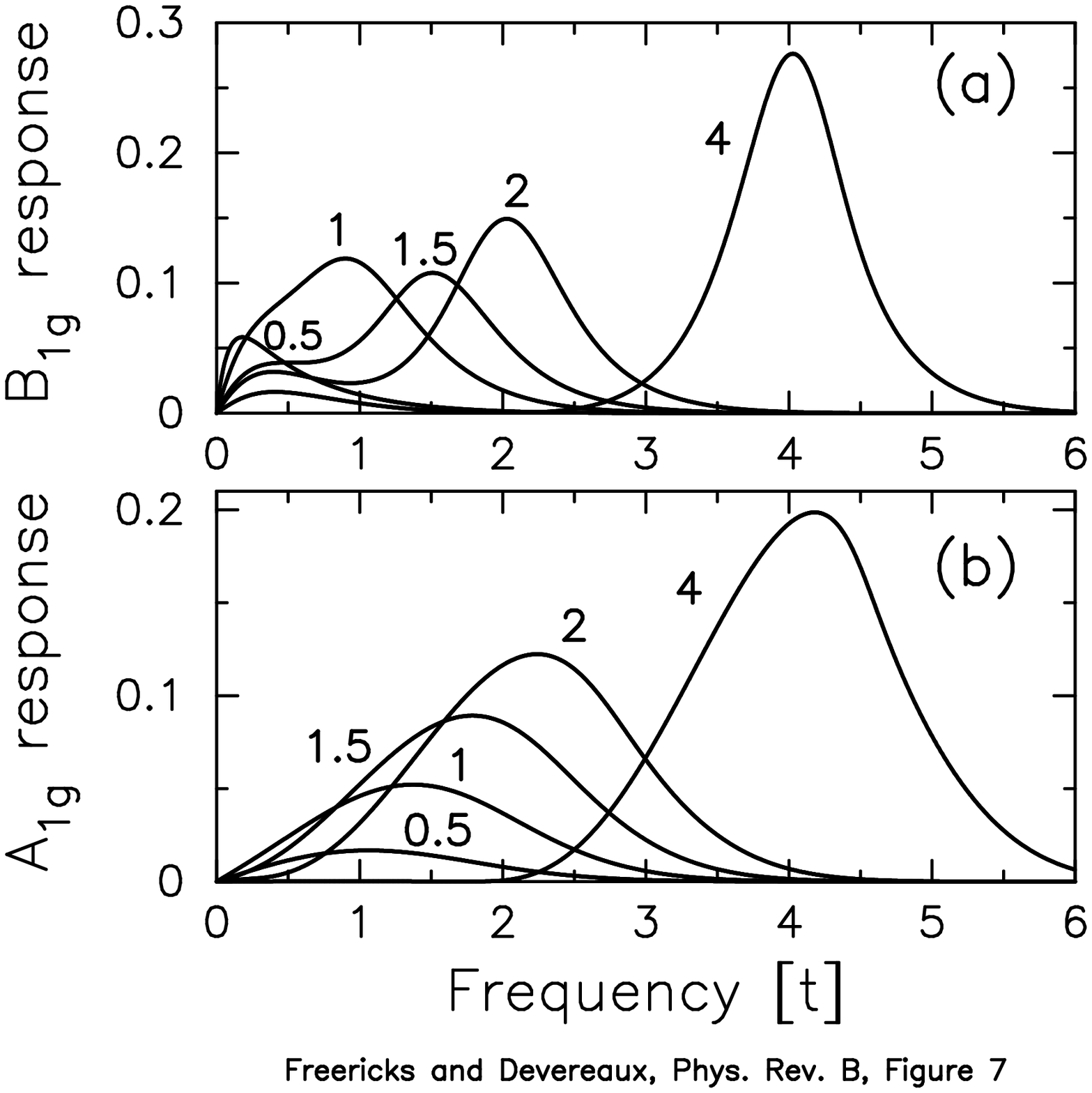}}
\vspace{.5cm}
\caption[]{Nonresonant (a) $B_{\rm 1g}$ and (b) $A_{\rm 1g}$
Raman response for different values of $U$ at $T=0.5$. The Raman response is 
measured in arbitrary units.  Notice how the vertex corrections suppress the 
low-frequency spectral weight in the $A_{\rm 1g}$ channel.
}
\label{fig: u}
\end{figure}   

In Figure \ref{fig: u}(a) we plot the nonresonant $B_{\rm 1g}$
Raman response at a fixed temperature
$T=0.5$ for different values of $U$. For small values of $U$,
a small scattering intensity is observed due to the weak interaction
among ``quasiparticles'' providing a small region of phase space allowable for
pair scattering. The peak of the response reflects the dominant
energy scale for scattering, as is well known in metals\cite{metals}
and the high-energy tail is the cutoff determined by the finite energy band.
This shape is also understandable from the Shastry-Shraiman relation---since
the optical conductivity is a Lorentzian, the Raman response is just
proportional to $a\nu/[\nu^2+a^2]$, which assumes the above form.
As $U$ increases, the low-frequency response is depleted as
spectral weight gets shifted into a large charge transfer peak at
a frequency $\sim U$. The charge transfer peak begins to appear for
values of $U$ for which the DOS is still finite at the Fermi level $(U=1)$
and becomes large in this pseudogap phase before growing even larger in
the insulating phase. Notice how low-frequency spectral weight remains
even as one is well on the insulating side of the quantum critical point
$(U=4)$ and at a temperature $T$ much lower than the gap.  It
is these spectral features that are characteristically seen in the experiments
and which can only be seen in a theory that approaches the quantum critical
point.     

In Fig.~\ref{fig: u}(b), we show the nonresonant $A_{\rm 1g}$ Raman response.
As shown above, the $A_{\rm 1g}$ response is independent of the value of
$c$.  The general behavior is similar to that
of the $B_{\rm 1g}$ channel except (i) at weak coupling the Raman scattering
is more symmetric and pushed to higher energy; (ii) the vertex corrections
suppress all nontrivial low-frequency Raman response; and (iii) the widths of
the charge-transfer peaks are enhanced.
 
\begin{figure}
\vspace{5mm}
\epsfxsize=3.0in
\centerline{\epsffile{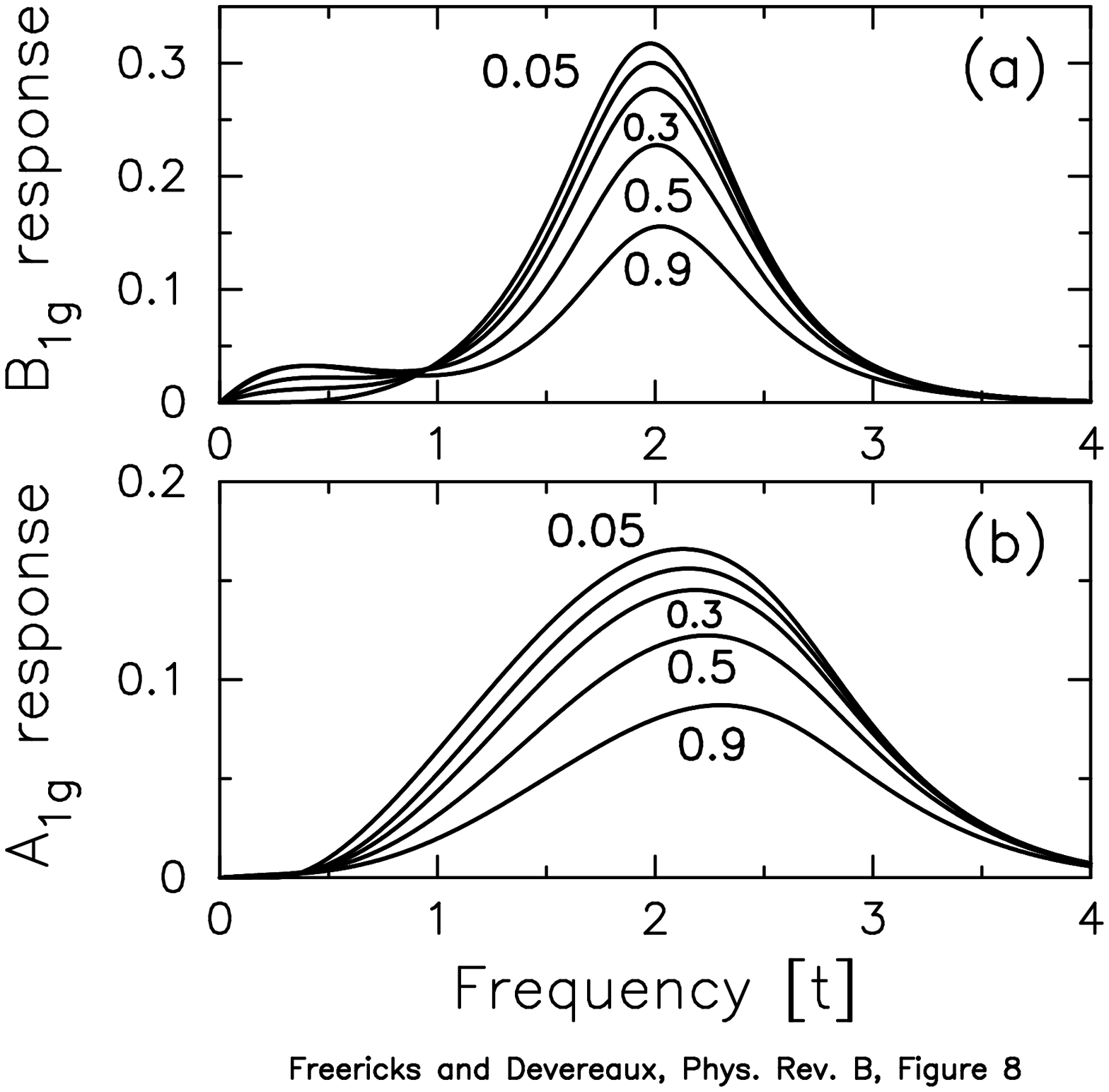}}
\vspace{.5cm}
\caption[]{(a) Nonresonant $B_{\rm 1g}$
Raman response for a range of temperatures ($T=0.05,0.2,0.3,0.5,0.9$) for $U=2$
(which lies just on the insulating side of the metal-insulator transition) and
(b) nonresonant $A_{\rm 1g}$
Raman response for the same temperatures at $U=2$. The lines are labeled by
their temperature (except for $T=0.2$ which is unlabeled).  Note how the
$B_{\rm 1g}$ response has low-frequency spectral weight that develops 
rapidly at an onset temperature of $T\approx 0.2$ (the low frequency response
at T=0.5 and T=0.9 overlap) and note the isosbestic
point at $\nu\approx 1$.  The ratio of the range in frequency over which
the low-frequency weight increases and the onset temperature is about 5.
There are no anomalous features in the $A_{\rm 1g}$ spectrum.
}
\label{fig: t}
\end{figure}   

Since the Raman response displays anomalous features on the insulating side of 
the metal-insulator transition, we first present results for $U=2$, just
on the insulating side of the quantum critical point.
In Figure \ref{fig: t}(a), we plot the temperature dependence.
The total spectral weight increases dramatically
with decreasing temperature as charge transfer processes become
more sharply defined. At the same time, the low-frequency response
depletes with lowering temperatures, vanishing at a temperature which is
on the order of the $T=0$ insulating gap (we are unable to analytically
estimate the crossover temperature). This behavior is precisely what
is seen in experiments on\cite{SLC2} FeSi and on\cite{irwin} underdoped
La$_{2-x}$Sr$_{x}$CuO$_{4}$ at low temperatures where both
the isosbestic point and the low temperature spectral weight depletion
can be seen.  Similar results are also seen\cite{SLC1} in SmB$_6$, but
a low-energy peak also develops in that material at low temperatures.
In the $A_{\rm 1g}$ channel [Fig.~\ref{fig: t}(b)], we see the
same sharpening of the charge-transfer peak at low $T$, but the low-energy
response is much smaller and changes much more slowly with $T$ (but, in fact,
increases as $T$ is lowered).

If one were to interpret the temperature at which the $B_{\rm 1g}$
Raman spectral weight starts to deplete as the ``transition temperature''
$T_c$ and the range of frequency over which the weight is depleted as the gap
$\Delta$, then one would conclude that near the quantum critical point $2\Delta/
k_BT_c \gg 1$. This is because the ``$T_c$'' is effectively determined by the 
gap in the single-particle density of states (which is small near the
quantum critical point), while the ``$\Delta$'' is
determined by the width of the lower Hubbard band (which remains finite at
the quantum critical point); hence the ratio can become
very large near the quantum critical point (and should decrease in the
large-$U$ limit).

\begin{figure}
\epsfxsize=3.0in
\centerline{\epsffile{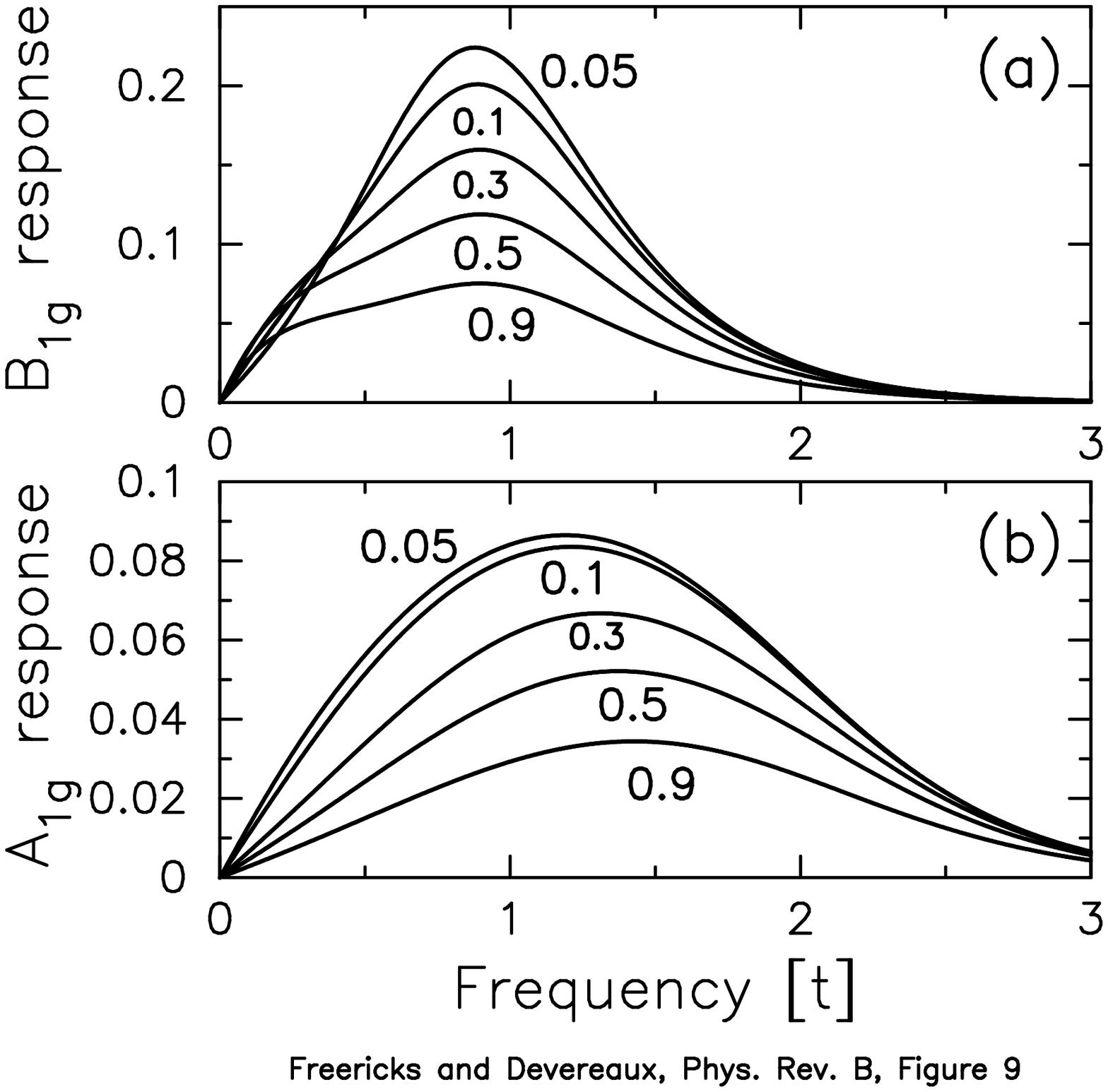}}
\caption{Nonresonant (a) $B_{\rm 1g}$ and (b) $A_{\rm 1g}$ Raman response
at $U=1$ for $T=$0.9, 0.5, 0.3, 0.1, 0.05. The lines are labeled by the
temperature.  Note how the $B_{\rm 1g}$ response develops an isosbestic
point at low temperatures, but the low-frequency depletion is more modest
here than in the insulating phase.  The $A_{\rm 1g}$ response has no
anomalous features, but the peak of the response moves to lower energies
as the Raman response sharpens at low temperature.  In the moderate-to-low
frequency range, the Raman response increases as $T$ is lowered.
\label{fig: u=1}}
\end{figure}

The nonresonant Raman response is plotted in Fig.~\ref{fig: u=1} for a number
of different temperatures at $U=1$.  Note how the $B_{\rm 1g}$ response has
nontrivial low-energy spectral weight, even though it is not completely
separated from the charge-transfer processes.  Even in this case, one can see
the low-temperature development of an isosbestic point near $U/2$ for
temperatures below about $T=0.3$.  As the low-energy spectral weight is
depleted, the peak becomes more symmetric in shape.  In the $A_{\rm 1g}$
channel, the response sharpens, and the peak moves to lower energy as the
temperature is lowered.  In fact, the low-energy spectral weight actually
increases as $T$ is lowered.  There is no indication of a separation of
the response into low and high-energy features that have a different
temperature dependence (as seen for the $B_{\rm 1g}$ response).

\begin{figure}
\epsfxsize=3.0in
\centerline{\epsffile{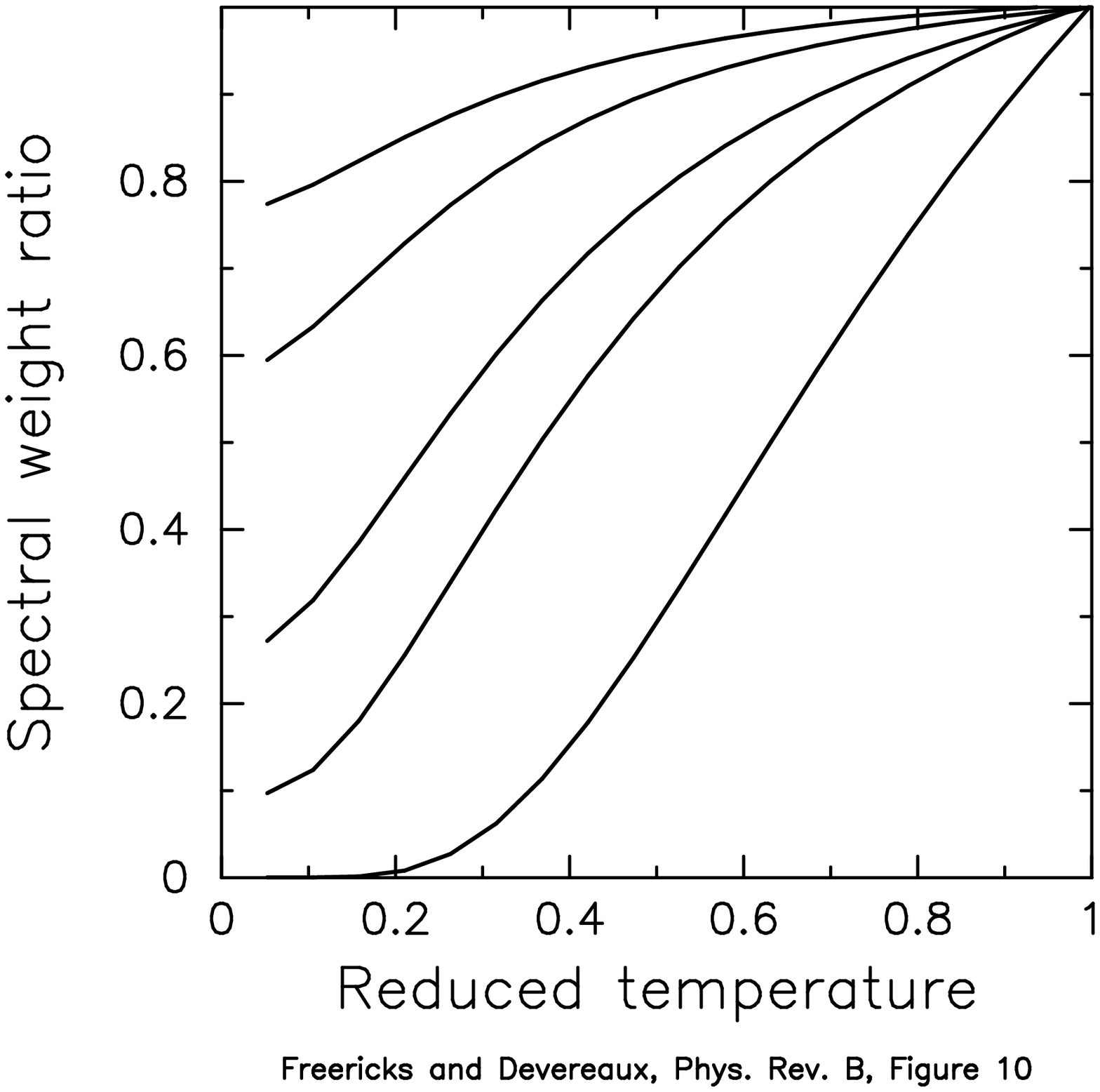}}
\caption{Ratio of low-frequency spectral weight at temperature $T$ to
the low-frequency spectral weight at $T=0.95$ plotted versus reduced temperature
$T/0.95$ for different values of $U$. The values of $U$ are 0.5,1,1.5,2, and
4, and the curves correspond to in creasing values of $U$ starting at
the top with 0.5 and running to the bottom at 4.
Note how the sharpening of the Raman response as $T$ is lowered results
in significant reductions to the low-frequency spectral weight even in the
metallic case.  In the strong insulator phase, the spectral weight 
decreases by over three orders of magnitude $(U=4$).
\label{fig: spectral_weight}}
\end{figure}

The $B_{\rm 1g}$ spectral-weight transfer from low frequencies to the charge
transfer peak as a function of temperature can be quantified by
separating the Raman response into two regions determined by the isosbestic
point and plotting the ratio of the
low-frequency spectral weight at temperature $T$ to the low-frequency
spectral weight at $T=0.95$  versus reduced temperature $T/0.95$
[Fig.~\ref{fig: spectral_weight}].
Choosing $U/2$ as the location of the isosbestic point (which divides the 
low-frequency and high-frequency regions), we find
that the reduction of spectral weight from high to low
temperatures is over 50 percent even in the weak pseudogap phase, and
decreases by well over three orders
of magnitude as $U$ increases into the insulating phase ($U=4$).

The origin of the isosbestic point in the nonresonant $B_{\rm 1g}$ response
is mysterious, but can be motivated by
the Shastry-Shraiman relation\cite{ss}.  Since the optical conductivity
satisfies a sum rule, the appearance of an isosbestic point there is not
surprising, as any decrease in low frequency spectral weight must be
compensated by a corresponding increase in high-frequency spectral weight,
so one might expect there to be a frequency where the response doesn't depend 
on $T$ (of course this doesn't establish that an isosbestic point must
exist, it just motivates such an existence).  The isosbestic point in the 
$B_{\rm 1g}$ Raman response then follows from the Shastry-Shraiman relation,
since multiplying the optical conductivity
by a frequency will not modify the appearance of an
isosbestic point.

We attribute the presence of anomalous low-frequency  and low-temperature
response in a system which is a strongly correlated insulator
to the appearance of thermally activated transport channels (indeed, the only
temperature dependence to the Raman response comes from Fermi factors in
an integral).
In the insulating phase at zero temperature, the only available intermediate
states created by the light must involve double site occupancy of a conduction
and a localized electron, with an energy cost of $U$. This gives the large
charge transfer peak. As the temperature is increased,
double occupancy can occur and as a result light can
scatter electrons to hop between adjacent unoccupied states either directly
or via virtual double occupancies. The number of electrons
which can scatter in this fashion increases with increasing temperature,
leading to an increase in the low-frequency spectral weight. The frequency
range for this low-frequency Raman response is determined by the lower
Hubbard bandwidth, which is typically much larger than the temperature
at which these features first appear. 

There is also a more mathematical explanation to the low-frequency spectral
response. If we examine
the integral for the B$_{\rm 1g}$ Raman response in Eq.~(\ref{eq: b1greal}),
we note three important points (i) the imaginary part of the Raman response
is proportional to the real part of the integrand [the terms within the 
braces in Eq.~(\ref{eq: b1greal})]; (ii) the integrand vanishes
if the Green function (and self-energy) at $\omega$ and $\omega+\nu$
are both real; (iii) all temperature
dependence arises from the Fermi factors, since both $G$ and $\Sigma$ are
temperature-independent. In the insulating regime, the DOS breaks into two
pieces, a lower subband centered at $-U/2$ with a width of $O(1)$ and an upper
subband at $U/2$ with a width of $O(1)$.  The Green's functions are complex
only when the frequency argument lies within one of the bands.  Hence there
are two main contributions to the Raman response: (i) intraband processes,
where $\omega\approx -U/2$ or $U/2$ and $\nu\approx 1$; and (ii) interband
processes, where $\omega\approx -U/2$ and $\nu\approx U$.  The interband
processes, with $\nu\approx U$ are what give rise to the charge-transfer peaks
seen in the Raman response; these processes survive even at $T=0$.  The 
intraband processes, with $\nu\approx 1$,
give rise to the low-frequency spectral features.  At low
temperatures, these features are proportional to $f(\omega)-
f(\omega+\nu)$ which can be approximated by $\exp(-U/2T)[1-\exp(\nu/T)]$ in
the insulating phase. 
Hence, we expect the low-frequency spectral weight to be proportional to 
$\exp(-U/2T)$ in the large-$U$ limit.   In the A$_{\rm 1g}$ channel, the 
charge vertex makes the
integrand more complicated to analyze (but the response still separates into
interband and intraband processes), and the vertex corrections end up
ultimately suppressing the low-frequency response.

\begin{figure}
\epsfxsize=3.0in
\centerline{\epsffile{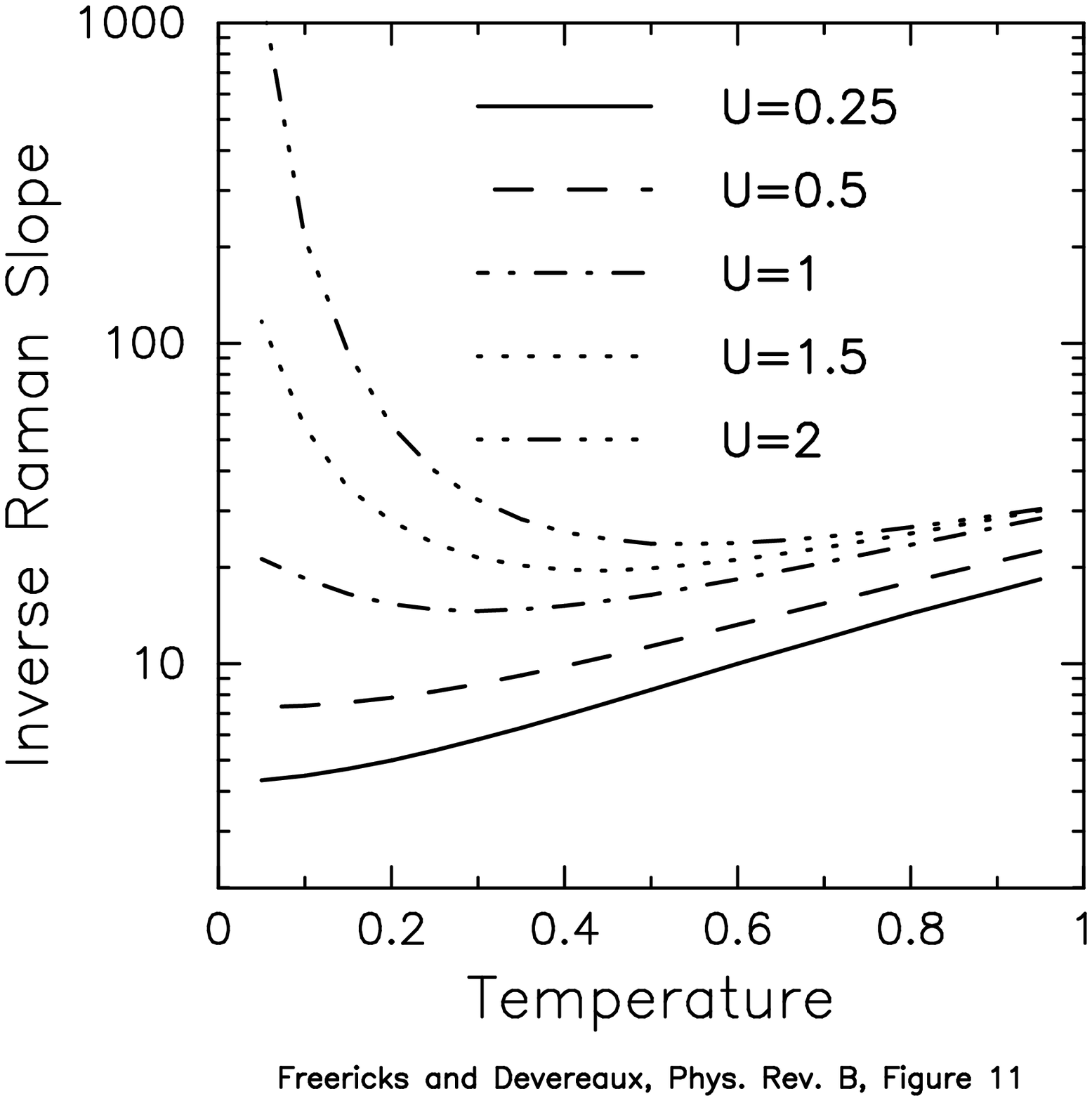}}
\caption{Inverse Raman slope as a function of temperature.  The inverse
Raman slope measures the scattering rate of the ``quasiparticles''
of a correlated metal.  As the system enters the pseudogap phase, the
inverse slope starts to increase at low temperatures, increasing
dramatically as one enters the correlated insulator $(U=2)$.
\label{fig: inv_raman_slope}}
\end{figure}

Lastly, we plot the inverse slope of the Raman response in 
Fig.~\ref{fig: inv_raman_slope}
as a function of temperature for different values of $U$. The inverse
Raman slope is the reciprocal of the slope of the Raman response in the
limit as $\omega\rightarrow 0$.
Since the self energy is
temperature independent, we might expect a constant Raman
slope as a function of temperature, as is the case
with a disordered Fermi liquid. However, this is not the
case due to the formation of a thermally generated band
for scattering.  For small values of $U$, the
temperature dependence of the Raman
inverse slope is weak due to the temperature independence of the
self energy. However, as the single-particle bands begin to separate,
the relevance of thermally generated
double occupancies becomes more pronounced and the inverse slope
rapidly rises at low temperatures. As the system becomes more insulating,
the low temperature inverse slope increases dramatically
due to the depletion of the low-frequency spectral weight.
As $U$ increases from the pseudogap phase into the insulating phase,
the temperature dependence of the Raman inverse slope indicates
the formation of gapped excitations (assuming the form
$T[1+\cosh (\Delta/\{2T\})]$, with $\Delta$ the gap in the single-particle
DOS).  Such behavior has been seen in the underdoped cuprate materials.

\section{Conclusions}

The electronic Raman response of a wide variety of correlated materials
(on the insulating side of the metal-insulator transition) display 
similar anomalous features, which point to a common explanation.  In our
results, we have shown how to see these anomalies in the $B_{\rm 1g}$
channel, by solving the Falicov-Kimball model. We saw that the Raman 
response ultimately depended only on the single-particle density of states.
In more complicated correlated models,
the metallic single-particle density of states can have an additional
Fermi liquid peak at
low frequencies which will add new features to the Raman 
response,\cite{hubbard_us} but on
the insulating side of the transition, where most of the anomalous
behavior is seen, the single-particle density of states must be
similar to that of the Falicov-Kimball model (except for some additional
weak temperature dependence of the interacting DOS), since an insulator has no
low-energy spectral weight. Hence, these anomalous Raman scattering results
are expected to be essentially
model-independent (since they only depend on the interacting DOS)! 

In this work, we also illustrated how one can calculate both the resonant and
the mixed Raman scattering response as well.  We showed how the bare diagrams
are renormalized for the different symmetry sectors, but did not perform
any numerical calculations here.

Our theoretical results compare quite favorably to the experimental results
seen in a variety of different materials ranging from mixed-valence
compounds \cite{SLC1}, to Kondo insulators \cite{SLC2} to the 
underdoped high-temperature 
superconducting oxides \cite{irwin,hackl,uiuc}.   
The experimental data
illustrate the three characteristic features seen in our theory: (i) there
is a rapid rise in the low-frequency spectral weight at low temperatures (at
the expense of the high-frequency spectral weight),
(ii) there is an isosbestic point, and (iii) the range of frequency over which
the low-frequency weight appears is much larger than the onset temperature,
where it is first seen.  Our model always produces an isotropic gap,
so we are unable to illustrate the symmetry selective behavior seen in the 
copper-oxides
where only the $B_{1g}$ response is anomalous, and the $A_{1g}$ and $B_{2g}$
responses are metallic.  But our results do indicate
a ``universality'' and model independence of the Raman response on the 
insulating side of, but in close proximity to, a quantum critical point.
We believe this is the reason why so many different materials show the
same generic behavior in their electronic Raman scattering.  

In future work we will examine nonresonant $B_{\rm 1g}$ Raman scattering in the
Hubbard model\cite{hubbard_us} and will examine resonant and mixed
Raman scattering effects in the Falicov-Kimball model.

\acknowledgments

J.K.F. acknowledges support of the National Science Foundation under grant 
DMR-9973225.  T.P.D. acknowledges support from the National Research and
Engineering Council of Canada. 
We also acknowledge useful discussions with S.L. Cooper, R. Hackl,
J.C. Irwin, M.V. Klein, P. Miller, A. Shvaika, and V. Zlati\'c. We also thank 
S.L. Cooper, R. Hackl, and J.C. Irwin for sharing their data with us.
J.K.F. thanks the hospitality of the IBM, Almaden Research Center, where this
work was completed.

\end{document}